# Near-Landauer-Bound Quantum Computing Using Single Spins

## Frank Z. Wang[1], SM, IEEE
[1] Division of Computing, Engineering & Mathematics Sciences, University of Kent, Canterbury, CT2 7NF, the UK

Corresponding author: Frank Z. Wang (email: frankwang@ieee.org).

This work was partially funded by an ERC grant, PIIFGA2012332059, Marie Curie Fellow: Prof. Leon Chua (UC Berkeley), Scientist-in-charge: Prof. Frank Wang (University of Kent).

**ABSTRACT** This study is the first experimental verification of Landauer's bound on a single spin, which is the smallest information carrier in size. We used four experiments (single spin experiment, giant spin experiment, nanomagnet experiment and Stern-Gerlach experiment) to demonstrate that a single spin was much more energy-efficient than other information carriers due to its small size and weak coupling with the surroundings. We conclude that quantum spintronics, with single spins as qubits, is an energy-efficient computing paradigm that requires the smallest amount of energy ($1.2 \times 10^{-26}\ J$ per spin qubit, close to the theoretical Landauer bound of $9.6 \times 10^{-27}\ J$ at $1\ mK$) to perform computations.

## I. INTRODUCTION

The engineering implementation of two-level quantum systems (qubits) matters in terms of building a working device. A quantum electronic device stimulates the interaction of coherent electromagnetic radiation on the quantum level and the transition between quantum energy levels.

Qubits have served as quantum counterparts to classical bits since the 1990s. As the unit of a classical computer, a classical bit can only take either the '0' or '1' state. To describe a classical bit sharply switching between '0' and '1', one must make some approximations to the initial Hamilton's equation (or its counterpart for an electric system). This is because, characterized by its position and momentum, a classical object can only take a value within a continuous range and continuously change with time [1].

A quantum state of a qubit with mutually orthogonal unit vectors $|0\rangle$ and $|1\rangle$ can be written as $|\Psi\rangle = \alpha|0\rangle + \beta|1\rangle$, where complex numbers $\alpha$ and $\beta$ satisfy the condition $|\alpha|^2 + |\beta|^2 = 1$ ($|\alpha|^2$ is the probability of finding the qubit in $|0\rangle$, and $|\beta|^2$ is the probability of finding the qubit in $|1\rangle$). To find a quantum system in a certain state, we need some physical quantities (observables). By measuring one of the observables many times, we will obtain the probability of finding the system in the corresponding state on average [1].

To date, many kinds of qubits have been successfully fabricated. One kind is charge qubits that use quantum dots (Figure 1). Such a dot is so small that it can contain no more than one electron at a given time since electrons with the same electric charge strongly repel each other. By putting one electron across two neighbouring quantum dots, we can actually obtain a charge qubit. The states $|0\rangle$ and $|1\rangle$ can be detected by measuring the local electric potential, and the electron can be moved around by applying a voltage with an approximate value [2].

If the dot (island) is made of a superconductor, then we obtain a superconducting charge qubit [Figure 2(a)], where $|0\rangle$ and $|1\rangle$ represent the presence or absence of a Cooper pair or BCS (Bardeen–Cooper–Schrieffer) pair of electrons [3], respectively.

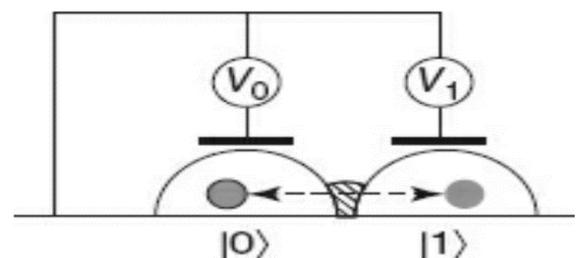

**FIGURE 1.** Charge qubit comprising two quantum dots [2].

Due to its special property (a superconducting loop can carry an electric current forever), the usage of a superconductor in a qubit makes it possible to protect its state from external noise. By interrupting a superconducting loop with a Josephson junction [4], we obtain a superconducting flux qubit ('flux' refers to the magnetic flux, produced by the current flow), as shown in Figure 2(b). The states $|0\rangle$ and $|1\rangle$ represent the counterclockwise current flow and the clockwise current flow, respectively.

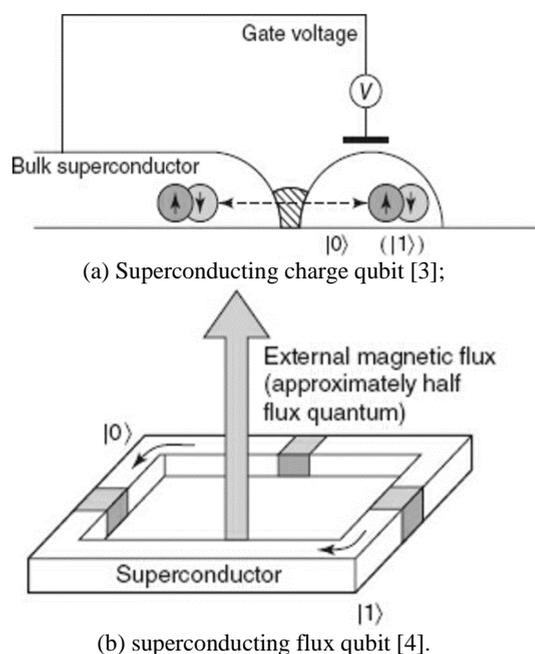

(a) Superconducting charge qubit [3];

(b) superconducting flux qubit [4].

**FIGURE 2.** Superconducting qubit can protect its state from external noise [3][4].

It is worth mentioning that the electric charge in the above charge qubit consists of one or two electrons [2][3], whereas the current flow in the above flux qubit involves (typically) more than two electrons [4]. Nevertheless, these qubits still behave like two-level quantum systems.

## II. SPINTRONICS FOR QUANTUM COMPUTING
In addition to the above discussed charge and flux qubits in conventional quantum electronics, another kind of qubit is the qubit based on spin, which is the angular momentum of a quantum particle, measured in units of $\hbar$: $L = \hbar s$ [5]. As an indispensable and inescapable property, the spin angular momentum cannot be reduced to zero or anything else. If a quantum particle as an information carrier is 'at rest' or 'anchored' (i.e., its kinetic energy has the minimal value allowed by Heisenberg's uncertainty principle [5]), then its spin is still present and can be manipulated and recorded. Scalably, the magnetic moment of a macroscopic particle consists of an integer number of spins. Such a macroscopic particle can also be manipulated by an external electromagnetic field. Being a quantum observable, spin is quantized such that $s$ can be either an integer or a half-integer: $s = \frac{1}{2}, \frac{3}{2}, ....$ In particular, an electron or a neutron has spin-1/2 and the moment carried by a proton is approximately 2.8 nuclear magneton [5].

If a 'resting' electron is trapped or anchored in a potential well, then its two spin states can play the role of a qubit. As shown in Figure 3, the difference in energy between the two states '0' (when the magnetic moment is parallel to the magnetic field $B$) and '1' (when it is antiparallel) is $\Delta E_{\uparrow\downarrow} = \mu_B B$.

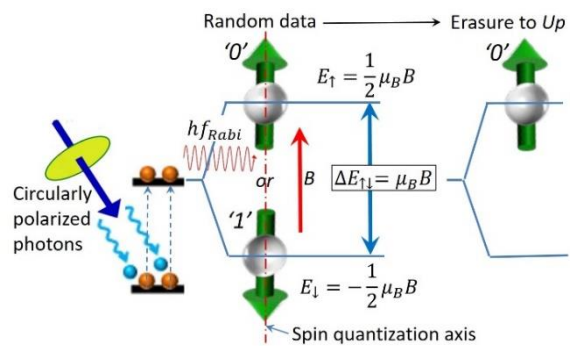

**FIGURE 3.** Erasure of a spin qubit from the random data ('0' or '1') state. A circularly polarized on-resonant $\mu m$ laser cyclically pumps the spin qubit to a defined quantum state, followed by spin reversal induced by either a magnetic field $B$ or Rabi flopping between the two levels illuminated with light exactly resonant with the transition occurs at the Rabi frequency.

The spin-related magnetic dipole moment of an electron is $\mu_B = 9.2740 \times 10^{-24} \ J/T$ (the Bohr magneton [5]), that is, very small. For a proton, $\mu_N = 5.0507 \times 10^{-27} \ J/T$ [5], that is, even smaller. Nevertheless, the weakness of the signal of each spin does not stop us from using spins as qubits.

One method is to use multiple identical molecules, each containing several atoms with nuclear spin-1/2 (Figure 4) – e.g. $^1H$, $^{13}C$ or $^{19}F$ [6]. These spins interact with each other and can be controlled and measured using the nuclear magnetic resonance (NMR) technique. The large number of spins involved compensates for the weakness of the signal of each spin.

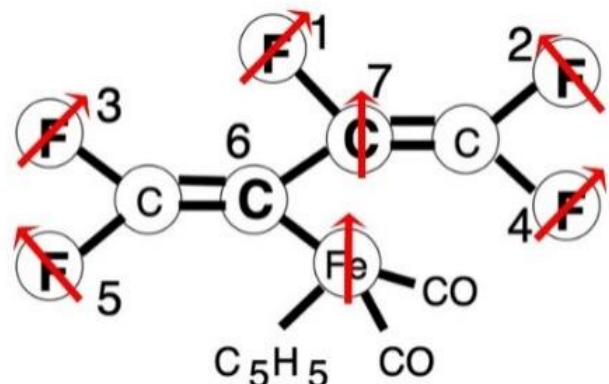

**FIGURE 4.** In 2001, an NMR quantum computer containing seven qubits (extracted from some atoms in a molecule) performed a real quantum algorithm (factoring the number 15 = 3 × 5) [6].

In 2014, a spin-spin magnetic interaction experiment [7] was carried out. A circularly polarized on-resonant laser (with a wavelength ranging from 422 $nm$ to 1092 $nm$) cyclically pumped two electrons bound within two ions across a separation ($d = 2.18 \sim 2.76\ \mu m$) to a well-defined quantum state $|\uparrow\downarrow\rangle$ or $|\downarrow\uparrow\rangle$. As shown in Figure 5, the two ions were entangled to protect the two (entangled) spins against errors (decoherence processes). Relevant to our study here, a single spin can be reliably switched with a typical detection fidelity of 98% [7].

Due to the spin-orbit coupling, a (weak) equivalent magnetic field acts on each electron in its rest frame (top right), i.e., bound electrons rather than free electrons are being used here. As the smallest magnet (the Bohr magneton), a spin ($1\ \mu_B$) applies a tiny magnetic field $[B_{spin-spin} = \frac{\mu_0}{4\pi}\frac{2\mu_B}{d^3} = (0.88 \sim 1.79) \times 10^{-13}\quad T$, where $\mu_0 = 4\pi \times 10^{-7} T \cdot m/A$ is the vacuum permeability constant] to another spin. That is, the measured magnetic interaction was extremely weak, six orders of magnitude smaller than magnetic noise (typically 0.1 $\mu T$).

The two-spin Hamiltonian in the above spin-spin experiment can be written as:

$$H = \underbrace{0.5\hbar(\omega_{A,1}\sigma_{z,1} + \omega_{A,2}\sigma_{z,2})}_{\text{Zeeman shift due to B (MHz)}} \underbrace{+ 2\hbar\zeta\sigma_{z,1}\sigma_{z,2} - \hbar\zeta(\sigma_{x,1}\sigma_{x,2} + \sigma_{y,1}\sigma_{y,2})}_{\text{Spin-spin magnetic interaction (mHz)}}\ (1)$$

Here $\sigma_{j,i}$ is the $j \in \{x, y, z\}$ Pauli spin operator of the $i$th spin; $\omega_{A,i} = 2\mu_B B_i/2\hbar$, where $B_i$ is the aforementioned external magnetic field, including the magnetic noise fluctuation at $kHz$. The spin interaction strength is $\zeta = \mu_0\mu_B^2/4\pi\hbar d^3$. As illustrated in the two spins' energy diagram in Figure 5 (middle), the first term on the right-hand side of Equation (1) describes the Zeeman shift of the spins' energy due to the external magnetic field $\boldsymbol{B}$, which is equivalent to $MHz$ in the spin Larmor frequency $\omega_{A,i}$ ($i = 1,2$) [7] that characterizes the precession of a transverse magnetization about a static magnetic field. As also illustrated in Figure 5 (middle), the second and third terms describe the weak, millihertz-scale spin-spin magnetic interaction that is insusceptible to spatially homogeneous magnetic noise [7]. In this study, it is the first term ($MHz$) that was our focus since its contribution to the energy consumption is nine orders of magnitude larger than the second and third terms due to the spin-spin magnetic interaction ($mHz$).

To introduce modulation in this spin-spin experiment [7], a spin rotation can be performed by pulsing a resonant oscillating magnetic field, perpendicular to the spin quantization axis, resulting in a Rabi frequency $f_{Rabi}$ of 65.8 $kHz$, as illustrated in Figure 3 [7].

Although this experiment was designed for a different purpose (the spin-spin magnetic interaction measurement), it equivalently incorporates the total work involved in a complete erasure protocol for a single spin. According to our calculations in this paper, this spin-spin experiment [7] is the first experimental validation of Landauer's bound on a single spin that is the smallest among various information carriers.

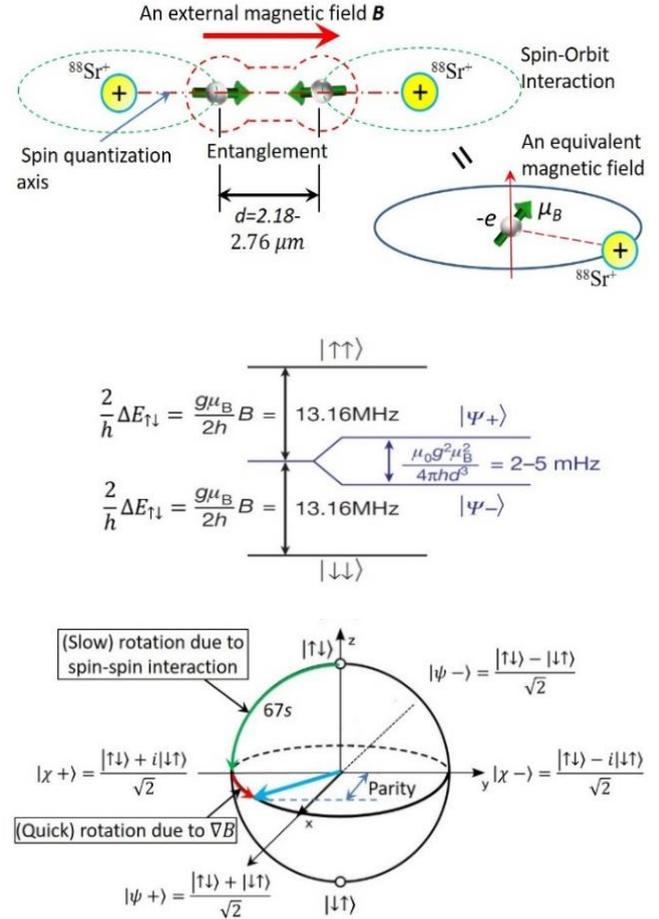

**FIGURE 5. Two outermost ground-state spin-1/2 valence electrons orbiting around two $^{88}$Sr+ ions [7]. To set the spin quantization axis (along which the angular moment has well-determined values), an external magnetic field $\boldsymbol{B}$ of $(1.3 \times 10^{-3} \pm 1.0 \times 10^{-7})$ $T$ is applied along the line linking the two electrons to induce spin rotation [7]. Spin–spin magnetic interaction induces $x$ rotation (green arc in the Bloch sphere in the bottom) from the north pole ($|\uparrow\downarrow\rangle$), through the fully entangled state ($|\chi+\rangle = (|\uparrow\downarrow\rangle + i|\downarrow\uparrow\rangle)/\sqrt{2}$) and towards the south pole ($|\downarrow\uparrow\rangle$). A (controlled) magnetic field gradient $\nabla B$ induces $z$ rotations (red arc in the Bloch sphere) so that $\boldsymbol{B}$ also facilitates the parity observable measurements (to measure the spin-spin magnetic interaction although it does not depend on $\boldsymbol{B}$ at all) on the equatorial plane after performing a collective π/2 spin rotation (green arc). The physical observable (the projection of the final Bloch vector on the $x$ axis) is first order sensitive to the spin-spin interaction strength and the experiment time ($T_{Bell} = 67$ s).**

In 2016, an energy dissipation of 4.2 zeptojoules was measured in a single-domain nanomagnet (comprising more than $10^4$ spins) [8] (Figure 6). The energy dissipation was calculated by the following equation

$$\int M \cdot dH = \int \{(M_x \times dH_x) + (M_y \times dH_y)\}$$
$$= \int M_x \times dH_x + \int M_y \times dH_y$$
$$= Area_{M_x - H_x\ loop} + Area_{M_y - H_y\ loop}.\quad (2)$$

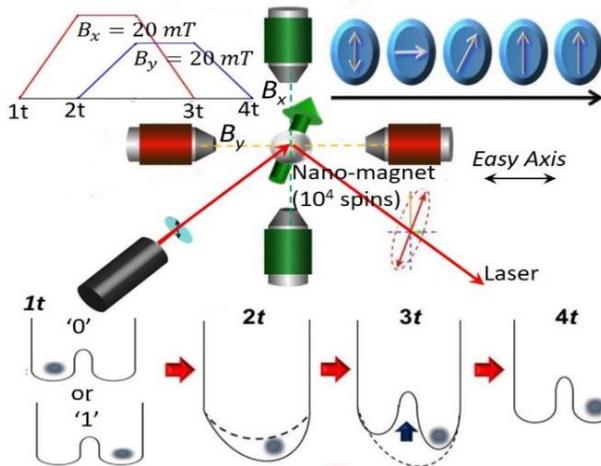

**FIGURE 6. Experimental nanomagnet bit (comprising $10^4$ spins) [8].** The magneto-optic Kerr microscopy was used to measure the energy dissipation. $B_x = 20\ mT$ is applied along the hard axis to remove the uniaxial anisotropy barrier (no tunnelling effect in this classical binary system), whereas $B_y = 20\ mT$ is applied along the easy axis to flip it to the erasure state ('1') at 3$t$.

In 2018, a giant spin reversal experiment [9] was reported, in which each nanomagnetic bit was composed of eight spin-$\frac{5}{2}$ $F_e^{3+}$ ions. These ions are coupled to each other by antiferromagnetic interactions (the spins of the ions Fe$_1$, Fe$_2$, Fe$_5$, Fe$_6$, Fe$_7$, Fe$_8$ are up and the spins of the Fe$_3$ and Fe$_4$ ions are down). A collective ($20\ \mu_B$) giant spin is then formed. With aligned magnetic axes, arrays of these molecular magnets can be packed into a single crystal (Figure 7, top). The Hamiltonian of this giant spin is

$$\mathcal{H} = -D_{ani}S_z^2 + E_{ani}(S_x^2 - S_y^2) - 2\mu_B \mathbf{S} \cdot \mathbf{B}, \quad (3)$$

in which the first two terms account for the magnetic anisotropy energy involving the two anisotropy constants ($D_{ani}/k_B = -0.294\ K$ and $E_{ani}/k_B = 0.046\ k$ where $k_B$ is the Boltzmann constant), and the third (Zeeman) term accounts for the interaction between the total spin $S$ and a magnetic field $B$ [9]. As shown in Figure 7 (bottom), an effective energy barrier was accordingly created, separating the $S_z = \pm 10$ ground eigenstates that encode the '↑' and '↓' bit states.

In this giant spin reversal experiment, two external magnetic fields ($H_y$, $H_z$) provide external control over the potential energy landscape. $\mu_0 H_y = (0\sim 2)\ T$ is applied along the medium axis, allowing one to tune the height of the potential energy barrier without breaking the degeneracy between '↑' and '↓' (Figure 7, bottom left). This transverse magnetic field promotes quantum mixing of the '↑' and '↓' spin orientations in this quantum system [9]. The giant spin can then tunnel through the barrier via progressively lower-lying energy levels, thus leading to spin reversal. In contrast, the other magnetic field $\mu_0 H_z = (0\sim 0.2)\ T$, parallel to the easy axis, favours either of the two eigenstates $S_z = \pm 10$ (i.e., by increasing the '↑' or '↓' polarization it selects), thereby adjusting the energy bias $\varepsilon$ (energy difference) between the two states (Figure 7, bottom right).

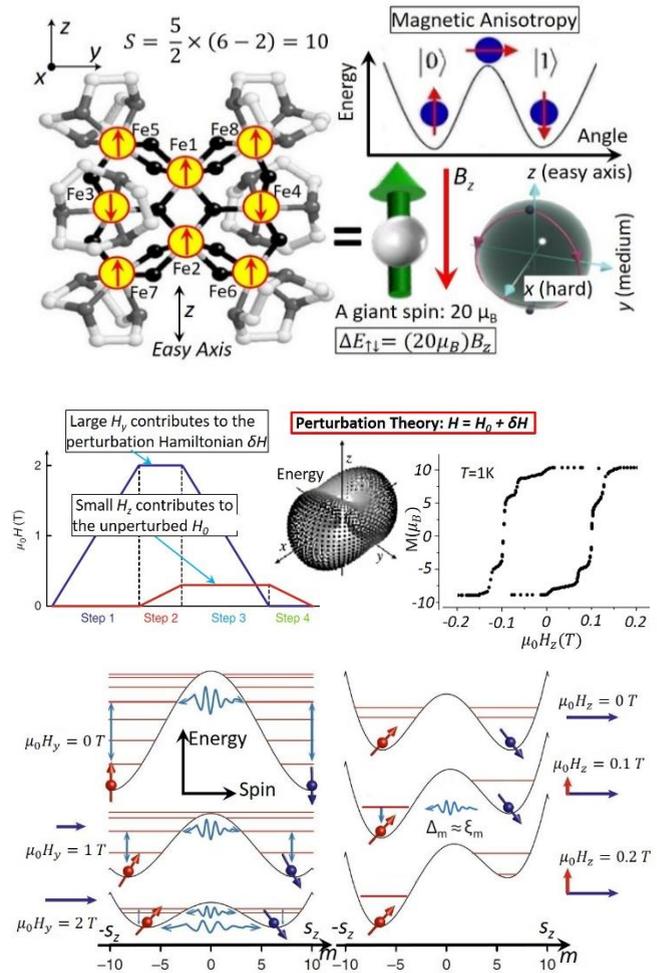

**FIGURE 7. A giant spin $S_z = \pm 10(20\ \mu_B)$ reversal experiment [9].** Here, **x, y** and **z** are defined as the hard, medium and easy magnetic axes, respectively. The (strong) magnetic anisotropy related to the local structure is illustrated (top right). Four-step sequence of magnetic fields $H_y$ (blue) and $H_z$ (red) induces the erasure process (middle left): in step 1, the magnetic field $\mu_0 H_y$ is ramped up to 2 $T$; in step 2 when $\mu_0 H_y$ is at its maximum, $\mu_0 H_z$ is ramped up to 0.2 $T$ to erase all molecular spins to the 'up' state; in steps 3 and 4, both $\mu_0 H_y$ and $\mu_0 H_z$ return to zero, completing the whole erasure process. The distance of the surface from the origin represents the classical potential energy of a spin (middle middle). A typical stepped hysteresis loop along the easy axis is shown (middle right). The transverse magnetic field $H_y$ lowers the barrier (bottom left), playing an assisting role (middle left). Although $|H_z| \ll |H_y|$, it is the horizontal magnetic field $H_z$ that contributes mainly to the total work of switching the giant spin via quantum spin tunnelling (bottom right).

It is worth highlighting here that, compared to the weak magnetic interaction ($\sim 10^{-13}\ T$) in the previous spin-spin experiment [7], the strength ($\sim 0.1\ T$) of the horizontal magnetic field $\mu_0 H_z$ required to flip this ($20\ \mu_B$) giant spin is much larger to overcome the strong magnetic anisotropy in the crystal lattice structure.

Figure 7 (bottom) illustrates spin relaxation combining tunnelling. The above process may take a long relaxation time, which can be improved by increasing $H_y$ to lower the energy level (bottom left). An energy gap $\Delta$ that equals the tunnelling bias $\varepsilon$ (to be defined in Figure 8) may trigger the quantum spin tunnelling for $E_{m=-S+1} = E_{m'=S}$ (bottom right, see details in Section IV).

There are two sides to everything. One problem with using multiple spins is their 'bulk' feature (all spins have to be controlled and measured at once) [1]. Another problem is that the strong coupling required to form a collective spin may result in disruption (from the environment) of the fragile quantum correlations essential for qubits [1]. A third problem is that some systems such as molecular nanomagnets exhibit a strong magnetic anisotropy that requires strong magnetic fields to drive them.

## III. SPIN QUBIT HAMILTONIAN AND SCHRÖDINGER EQUATION

The (time-independent) Hamiltonian of a spin qubit can be written in the Hermitian form with a special kind of symmetry (its diagonal elements being real and off-diagonal elements being complex conjugates) [1][5] as follows:

$$\hat{H} = -\frac{1}{2}\begin{pmatrix} \varepsilon & \Delta \\ \Delta & -\varepsilon \end{pmatrix}, \tag{4}$$

where $\varepsilon$ is the bias (shown in Figure 8), and $\Delta$ is the tunnelling splitting w.r.t. the ground state (to be shown in Figures 9 and 13).

Consider a ball that is rolling on an uneven surface [Figure 8(a)] and will eventually settle in one of the local minima, with the minimum potential energy $E_{min} = mass \cdot h_{min}$ ($h$ denotes the height of the ball). If two such minima are separated by a hill (an energy barrier), forming a two-well potential, then we can consider one of the two wells as '0' and the other as '1'. That is, the ball is our information carrier in a classical information system and its position is encoded as a binary bit of information. We may also assume that friction eventually stops the ball such that we can neglect its kinetic energy in this case (for a similar reason, we should use bound rather than free electrons while implementing quantum computing in spintronics). As shown in Figure 8(a), we can also create an energy difference $\varepsilon$ between the two states: $\varepsilon = E_1 - E_0$.

Once the ball is in '0', it will stay there forever, even if $E_1 < E_0$. To force the ball to move from '0' to '1', we have to push it over the hill (that is, Landauer's bound [10] in a classical information system; see the details in Section IV) by providing sufficient kinetic energy.

If the setup in Figure 8(a) is used for quantum computing (the ball is small enough and the barrier is narrow enough), then the ball can tunnel through the barrier with a rate $\Gamma$, as shown in Figure 8(b) [it can be considered as the quantum analogue of Figure 8(a)]. Approximately, this tunnelling process does not require any extra energy and resembles teleportation [1].

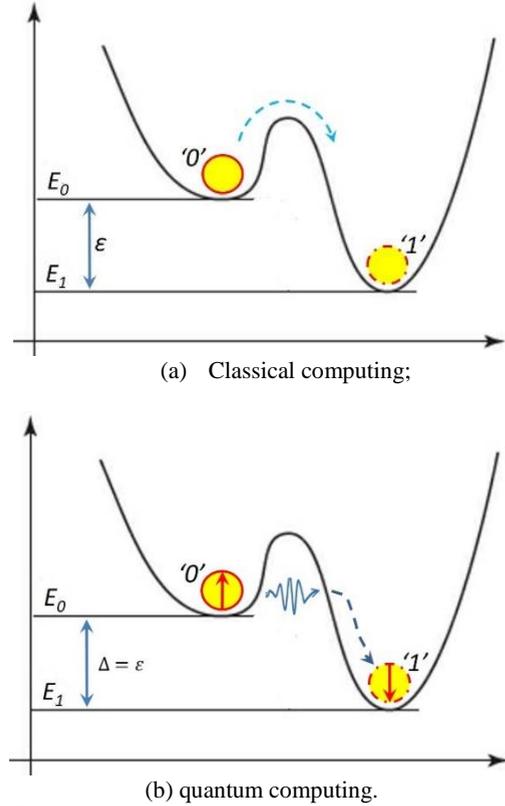

(a) Classical computing;

(b) quantum computing.

**FIGURE 8.** The two-well potential is used in both classical computing (a) and quantum computing (b). If the ball is small enough and the barrier is narrow enough, then the ball can tunnel through the hill with no need to climb it [1].

The spin qubit Hamiltonian in the Schrödinger equation is

$$i\hbar \frac{d}{dt}\begin{pmatrix} \alpha \\ \beta \end{pmatrix} = -\frac{1}{2}\begin{pmatrix} \varepsilon & \Delta \\ \Delta & -\varepsilon \end{pmatrix}\begin{pmatrix} \alpha \\ \beta \end{pmatrix} = \begin{pmatrix} -\frac{1}{2}\varepsilon\alpha & -\frac{1}{2}\Delta\beta \\ -\frac{1}{2}\Delta\alpha & -\frac{1}{2}\varepsilon\beta \end{pmatrix}, \tag{5}$$

where the 'energy' $\Delta$ is the tunnelling matrix element of the Hamiltonian [1].

This matrix equation is decomposed into a set of two equations: one corresponds the first row, and the other to the second row. If the qubit is not biased ($\varepsilon = 0$), then we obtain

$$i\hbar \frac{d\alpha}{dt} = -\frac{1}{2}\Delta\beta; \; i\hbar \frac{d\beta}{dt} = -\frac{1}{2}\Delta\alpha, \tag{6}$$

$$\frac{d\alpha}{dt} = \frac{i\Delta}{2\hbar}\beta; \; \frac{d\beta}{dt} = \frac{i\Delta}{2\hbar}\alpha. \tag{7}$$

Let us start from the state $|0\rangle$ ($\alpha = 1, \beta = 0$). We obtain $\frac{d\alpha}{dt} = 0$ and $\frac{d\beta}{dt} = \frac{i\Delta}{2\hbar}$, which means that the chance of finding our system in state $|1\rangle$ changes at a rate $\Gamma \sim \frac{\Delta}{\hbar}$. Note that $\Delta \sim \hbar\Gamma$, i.e., the Planck constant $\hbar$ times the frequency $\Gamma$ yields the 'energy' $\Delta$.

For a qubit, a solution (an energy eigenstate) in the form of $|\Psi_E\rangle = \begin{pmatrix} \alpha \\ \beta \end{pmatrix} exp\left(-\frac{iEt}{\hbar}\right)$ satisfies the stationary Schrödinger equation:

$$-\frac{1}{2}\begin{pmatrix}\varepsilon & \Delta \\ \Delta & -\varepsilon\end{pmatrix}\begin{pmatrix}\alpha \\ \beta\end{pmatrix} = E\begin{pmatrix}\alpha \\ \beta\end{pmatrix}, \quad (8)$$

where $E$ is a definite energy [1].

These two equations for two unknown quantities can be resolved only if

$$E = \pm\frac{1}{2}\sqrt{\varepsilon^2 + \Delta^2}, \quad (9)$$

in which the minus sign corresponds to the energy $E_g$ of the ground state $|\Psi_g\rangle$ and the plus sign to the energy $E_e$ of the excited state $|\Psi_e\rangle$, as shown in Figure 9.

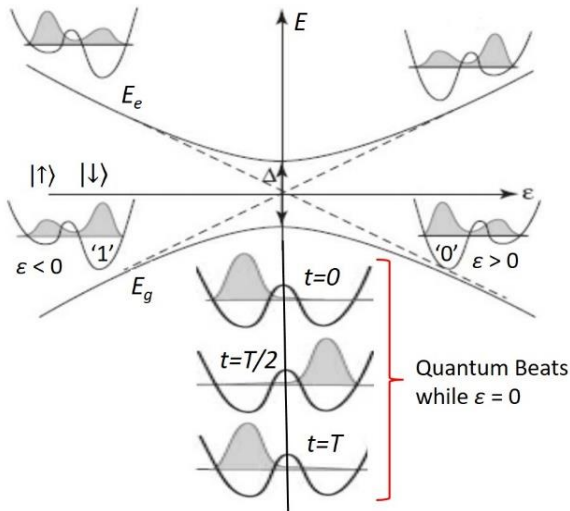

FIGURE 9. Energies and wavefunctions of the ground and excited spin qubit states. The state of the system is either '0' or '1' while $\varepsilon \neq 0$. Quantum beats [1][10] can be seen when $\varepsilon = 0$ [to be further elaborated in Figure 13(a)].

From the above equations, we obtain that

$|\Psi_g(\infty)\rangle = \begin{pmatrix}1\\0\end{pmatrix}$ and $|\Psi_e(\infty)\rangle = \begin{pmatrix}0\\1\end{pmatrix}$, when $\varepsilon \to \infty$;

$|\Psi_g(-\infty)\rangle = \begin{pmatrix}0\\1\end{pmatrix}$ and $|\Psi_g(-\infty)\rangle = \begin{pmatrix}1\\0\end{pmatrix}$, when $\varepsilon \to -\infty$;

$|\Psi_g(0)\rangle = \frac{1}{\sqrt{2}}\left[\begin{pmatrix}1\\0\end{pmatrix} + \begin{pmatrix}0\\1\end{pmatrix}\right]$ and

$|\Psi_e(0)\rangle = \frac{1}{\sqrt{2}}\left[\begin{pmatrix}1\\0\end{pmatrix} - \begin{pmatrix}0\\1\end{pmatrix}\right]$, when $\varepsilon = 0$.

The above relations imply that a large bias $\varepsilon$ (when the two wells are far apart) results in the state of the system being in either of the two wells [1], as shown in Figure 9. This phenomenon is important to ensure the irreversibility required in the information erasure (to be elaborated later).

At zero bias ($\varepsilon = 0$), the energies of the two wells coincide, but the energy gap between the ground and excited states is $\Delta$ (the tunnelling splitting). The eigenstates are superpositions of the '↑' and '↓' states:

$|\Psi_g(0,t)\rangle = \frac{1}{\sqrt{2}}[|\uparrow\rangle + |\downarrow\rangle]exp\left(\frac{i\Delta t}{2\hbar}\right)$ and $|\Psi_e(0,t)\rangle = \frac{1}{\sqrt{2}}[|\uparrow\rangle - |\downarrow\rangle]exp\left(-\frac{i\Delta t}{2\hbar}\right)$.

If the system is in state $|\uparrow\rangle$ at $t = 0$, then it can be written as

$$|\uparrow\rangle = \frac{1}{2}[(|\uparrow\rangle + |\downarrow\rangle) + (|\uparrow\rangle - |\downarrow\rangle)]$$
$$= \frac{1}{2}[\sqrt{2}|\Psi_g(0,t=0)\rangle + \sqrt{2}|\Psi_e(0,t=0)\rangle]. \quad (10)$$

To find the state of the system at any later time, we need to include the time-dependent vectors and obtain

$$|\Psi(t)\rangle = \frac{1}{2}[\sqrt{2}|\Psi_g(0,t)\rangle + \sqrt{2}|\Psi_e(0,t)\rangle]$$
$$= \frac{1}{2}\left[|\uparrow\rangle\left(e^{\frac{i\Delta t}{2\hbar}} + e^{-\frac{i\Delta t}{2\hbar}}\right) + |\downarrow\rangle\left(e^{\frac{i\Delta t}{2\hbar}} - e^{-\frac{i\Delta t}{2\hbar}}\right)\right]$$
$$= \left[|\uparrow\rangle\cos\frac{\Delta t}{2\hbar} + i|\downarrow\rangle\sin\frac{\Delta t}{2\hbar}\right]. \quad (11)$$

At time $t = \pi\hbar/\Delta t$, the system will be in the $|\downarrow\rangle$ state although we did nothing to it. It will keep oscillating between $|\uparrow\rangle$ and $|\downarrow\rangle$. The probabilities of finding the qubit in states $|\uparrow\rangle$ and $|\downarrow\rangle$ are

$$P_\uparrow(t) = |\alpha|^2 = |\langle\uparrow|\Psi(t)\rangle|^2 = \left(\cos\frac{\Delta t}{2\hbar}\right)^2 = \frac{1}{2}\left(1 + \cos\frac{\Delta t}{\hbar}\right),$$

and

$$P_\downarrow(t) = |\beta|^2 = |\langle\downarrow|\Psi(t)\rangle|^2 = \left(\sin\frac{\Delta t}{2\hbar}\right)^2 = \frac{1}{2}\left(1 - \cos\frac{\Delta t}{\hbar}\right),$$

respectively.

This leads to a phenomenon called quantum beats [1][10] (Figure 9) and the period is $\frac{2\pi}{\Gamma} = \frac{2\pi\hbar}{\Delta}$. The above deductions vividly demonstrate that the quantum dynamics of a spin qubit are largely determined by its energy time product in terms of setting the maximum speed at which a spin can modify its energy by a given amount. Further elaborations will be provided in the following sections.

On the other hand, the tunnelling in the spin-flipping by a magnetic field is irreversible since the magnetic field only favours flipping a spin with the opposite direction. This magnetic field tilts the potential landscape such that $\varepsilon \neq 0$, as shown in Figure 7 (bottom right), Figure 8(b) and Figure 13(b), respectively. Thus, a single spin can still be irreversibly switched without having the (quantum beats) problem that the information carrier in the destination well 'tunnels' back to the original well in the completed erasure. Such reliable, irreversible switch has been experimentally observed in both the spin-spin magnetic interaction experiment (with a typical detection fidelity of 98%) [7] and the (20 $\mu_B$) giant spin reversal experiment [9].

## IV. LANDAUER'S BOUND AND QUANTUM SPIN TUNNELLING

If we take the strongest instantaneous magnetic field of 1,200 Tesla created in the laboratory [12] as $B$ in Figure 3, then $\Delta E_{\uparrow\downarrow} = \mu_B B \approx 10^{-18} J$. Table 1 lists the energies to flip a spin with different magnetic fields in comparison to Landauer's bound [11].

In Section II, we mentioned that, in contrast to the weak external magnetic field $B$ of $1.3 \times 10^{-3}$ $T$ to erase a single spin $(\Delta E_{\uparrow\downarrow} = \mu_B B \approx 1.2 \times 10^{-26} \, J)$ in the spin-spin experiment [7], the strength ($\sim0.1 \, T$) of the horizontal magnetic field required to erase the ($20 \, \mu_B$) giant spin is extremely large to overcome the strong magnetic anisotropy in the crystal lattice structure [9]. As another comparison, the equivalent magnetic field (due to the spin-orbit coupling as shown in Figure 5) acting on the outermost electron orbiting around an $^{88}$Sr+ ion in the rest frame of this electron is only $\frac{z^4}{n^3} \approx 10^{-7} \, T$ [5], as listed in Table 1.

We take an average value for $B_z = \mu_0 H_z = 0.1T$ in the giant spin experiment because, at the end of Step 2 when $\mu_0 H_z$ is ramped up from $0 \, T$ to $0.2 \, T$ (Figure 5, middle left), the giant spins are already oriented to the 'up' configuration [9] (i.e., the erasure is already completed at this point).

TABLE 1. Energies to flip a spin with different magnetic fields (sorted by amplitude from large to small) in comparison to Landauer's bound ($k_B T \ln 2$) [11].

| Scenario, in which a spin or a giant spin is placed. | Magnetic Field $B$ | Energy ($\mu_B B$) to Flip a Spin in $B$ |
|---|---|---|
| Strongest magnetic field in the universe [12] | $10^9 \, T$ | $10^{-12} \, J$ |
| Strongest magnetic field in the lab [13] | 1,200 T | $10^{-18} \, J$ |
| Strongest (steady) magnetic field in the lab [13] | 45.22 T | $5 \times 10^{-20} \, J$ |
| **Landauer's bound [11] at 300 $K$** | **n/a** | **$10^{-21} \, J$** |
| Magnetic resonance imaging (MRI) [14] | 9.4 T | $10^{-22} \, J$ |
| (20 $\mu_B$) giant spin reversal experiment at 1 $K$ [9] | 0.1 T | $1.9 \times 10^{-23} \, J$ |
| **Landauer's bound [11] at 1 $K$** | **n/a** | **$10^{-23} \, J$** |
| Transformer [15] | 1 T | $10^{-23} \, J$ |
| External magnetic field $B$ in the spin-spin magnetic interaction at 1 $mK$ [7][this paper] | 1.3 $\times 10^{-3} T$ | $1.2 \times 10^{-26} \, J$ |
| **Landauer's bound [11] at 1 $mK$** | **n/a** | **$10^{-26} \, J$** |
| Earth's magnetic field [16] | $10^{-5} T$ | $10^{-28} \, J$ |
| Outermost electron orbiting around $^{88}$Sr+ in the spin-spin experiment [7][17] | $10^{-7} \, T$ | $10^{-30} \, J$ |
| Outermost electron in a Rydberg atom ($n$=630) [5] | $10^{-9} \, T$ | $10^{-32} \, J$ |
| Brain magnetic field [16] | 1 pT | $10^{-35} \, J$ |

In Section III, we mentioned that the tunnelling process approximately does not require any extra energy. However, it is only theoretically possible to use zero or an arbitrarily small energy to flip a spin via tunnelling, whereas in practice this energy must have a lower bound that is Landauer's bound. As listed in Table 1, the smallest amount of the energy of erasing an information bit is $1.2 \times 10^{-26} \, J$ with a single spin, which is already close to Landauer's bound. This is analogous to the world record of the lowest temperature ($3.8 \times 10^{-11} \, K$) [18], although in theory one can get as close as possible to absolute zero.

The so-called Landauer's bound was proposed by Rolf Landauer in 1961 [11], who argued that information is physical and that the erasure of a bit of classical information requires a minimum energy amounting to $\Delta E = k_B T \ln 2$, where $T$ is the temperature. Profoundly, Landauer's bound (Figure 10) is accepted as one of the fundamental limits in physics and computer science [19].

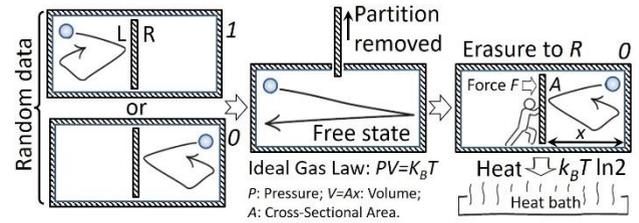

FIGURE 10. **For a bit of position-encoded classical information, the designated erasure state ('0') is reached from the random data state via the free state by removing the partition. The erasure dissipates the heat of $k_B T \ln 2$ (Landauer's bound [11]) by pushing it towards $R$ and exerting the work:** $W = \int_1^{0.5} Fd(1-x) = \int_1^{0.5} PAd(1-x) = -\int_1^{0.5} \frac{PAx}{x} dx = -\int_1^{0.5} \frac{k_B T}{x} dx = k_B T \ln 2.$ **If it is a bit of orientation-encoded classical information, then we still need to input a certain amount of energy above Landauer's bound to reach the designated erasure state from the random data state [8] since the orientation may fluctuate to take an arbitrary direction due to thermal agitation [17].**

Since 2012, Landauer's bound has been experimentally verified for various information carriers, including a single silica glass bead [20], a fluorescent particle [21], a single-domain nanomagnet [8], a single atom [22], a giant spin [9], and bacterial ion channels [23], as summarized in Table 2.

Specifically, the sizes of various information carriers and the experimental temperatures are listed in Table 2. Normally, quantum effects are observed at low temperatures and small sizes. For example, when the size of a magnetic particle decreases, it may be possible to invert the magnetization through the quantum tunnelling effect [9]. This effect should appear at low temperatures (when the spin is in its ground state), where it provides an energy-efficient path for magnetic relaxation only if the wavefunction of the left well overlaps with that of the right well.

TABLE 2. Experiments of Landauer's bound [11] on various information carriers (sorted by the size of the information carrier from large to small, except for the bacterial ion channels [23]).

| Year | Details (Information Carrier) | Size | $T$ ($K$) | Mode |
|---|---|---|---|---|
| 1961 | Landauer [11] | | | Classical |
| 2012 | Silica glass bead [20] | 2 $\mu m$ | 300 $K$ | Classical |
| 2014 | Fluorescent particle [21] | 200 $nm$ | 300 $K$ | Classical |
| 2016 | Single-domain nanomagnet [8] | $10^4$ spins | 300-400 $K$ | Classical |
| 2018 | Single-atom [22] | $^{40}Ca^+$ | 15 $\mu K$ | Quantum |
| 2018 | Giant spin [9] | 20 spins | 1 $K$ | Quantum |
| 2020 | Bacterial ion channels [23] | >> spins | | Classical |
| 2023 | Single spin [7][this paper] | 1 spin | 1 $mK$ | Quantum |

Noticeably, as listed in Table 1, the strength ($\sim0.1 \, T$) of the horizontal magnetic field in the giant spin reversal experiment [9] is 100 times the external magnetic field ($\sim10^{-3} \, T$) in the single spin experiment [7]. The detailed calculation follows.

As shown in Figure 7 (top right, bottom right), it is the horizontal magnetic field $H_z$ that contributes mainly to the total work of switching the giant spin via quantum spin tunnelling (in the same way as how we switch a single spin in Figure 3). As shown in Figure 7 (bottom left), the transverse magnetic field $H_y$ only plays an assisting role (lowering the barrier) on the switched giant spin via quantum spin tunnelling (although $|H_y| \gg |H_z|$). In theory, a transverse magnetic field is only used to remove the anisotropy energy barrier (Figure 6 and Figure 7) to facilitate a possible tunnel through the barrier. In practice, according to the giant spin reversal experiment [9], the measured work done by $H_y$ was $(-6 \pm 2) \times 10^{-24} J$ per molecule, much smaller than the measured work [$(1.7 \pm 0.3) \times 10^{-23} J$ per molecule] done by $H_z$.

Thus, the work required for erasure of each $(20\mu_B)$ giant spin qubit can be estimated by a simple formula:

$$\Delta E_{\uparrow\downarrow}^N = (N\mu_B)B_z \qquad (12)$$
$$= (20 \times 9.274 \times 10^{-24} J \cdot T^{-1}) \times 0.1 T$$
$$= 1.855 \times 10^{-23} J.$$

This estimation based on $B_z$ is at the same order of magnitude as the experimentally measured total work $(1.1 \pm 0.3) \times 10^{-23} J$ [9] and can be further adjusted by considering the perturbation brought by $B_y$ (to be elaborated later).

Similarly, the work required for erasure of each $(10^4$ spins$)$ nanomagnet qubit can be estimated by

$$\Delta E_{\uparrow\downarrow} = 10^4 \mu_B B_x + 10^4 \mu_B B_y \approx 4.0 \times 10^{-21} J, \qquad (13)$$

in which $B_x = 20 \, mT$ is applied along the hard axis to remove the uniaxial anisotropy barrier (no tunnelling effect in this classical binary system), whereas $B_y = 20 \, mT$ is applied along the easy axis to flip it to the erasure state ('1') at $3t$. Note that Equation (13) takes the same form as Equation (2) used in the nanomagnet experiment [8] in terms of the magnetization being expressed by the Bohr magneton in a single or collective spin and the magnetic fields taking their averages over the corresponding periods. This estimation perfectly matches the experimentally measured total work ($\sim 4 \times 10^{-21} J$) [8].

Equation (12) indicates that both the moment size $N$ of an information carrier and the applied magnetic field $B$ matter. This is important in terms of implementing quantum computing in spintronics, in which the formula $k_B T \ln 2$ introduced by Landauer in 1961 [11] may not be directly used for the following reason.

Equation (12) provides an analytical formula that is useful in practice to identify the dominant factor(s) in a four-step erasure process (Figure 7, middle left), in which each magnetic field has its ramping-up stage, constant (maximum) stage and ramping-down stage. Namely, this formula $\Delta E_{\uparrow\downarrow}^N = N\mu_B B_z$ considers only Step 2, 3 & 4 with $H_z$ [9] but it has taken into consideration the contributions in all four steps in the giant spin reversal experiment (Figure 7) and all contributions (including the external magnetic field to set the spin quantization axis and facilitate the parity observable measurements on the equatorial plane after performing a collective $\pi/2$ spin rotation, which we ignored in our previous work [17]) in the single spin experiment (Figure 5).

Note that the amplitude of $B_y = (0\sim2) \, T$ is 10 times that of $B_z = (0\sim0.2) \, T$, which implies that any hysteresis of $M_y$

[measured in Figure 11(b)] will be 'amplified' by a factor of 10 in terms of subtracting $M_y B_y$ from $M_z B_z$ [measured in Figure 11(a)] to obtain the net work. For this reason, although the measured work [$W_{1,3} = (-6 \pm 2) \times 10^{-24} J$] done by $H_y$ was namely 35% of the measured work [$W_{2,4} = (1.7 \pm 0.3) \times 10^{-23} J$] done by $H_z$ [9], the actual 'noise' level is only 3.5% compared to the original signal level, which is not uncommon.

Nevertheless, the concerned $H_y$ contribution to the total work in a complete cycle to erase the giant spin can still be estimated by following the Stoner–Wohlfarth (S-W) model [24] and the high-order perturbation theory [25] in the form of (the spin Hamiltonian $\mathcal{H} = \mathcal{H}_0 + \delta\mathcal{H}$, in which $\mathcal{H}_0$ is the unperturbed Hamiltonian that commutes with $S_z$ and $\delta\mathcal{H}$ is the perturbation Hamiltonian) [25]. Having rewritten Equation (3) as: $\mathcal{H} = (-D_{ani}S_z^2 - 2\mu_B S_z \cdot B_z) + (E_{ani}S_y^2 - 2\mu_B \boldsymbol{S_y} \cdot B_y) = \mathcal{H}_0 + \delta\mathcal{H}$, we found that, if a transverse magnetic field $(B_y \neq 0)$ is introduced, then $S_y^2 \neq 0$, resulting in the nonzero perturbation Hamiltonian $(\delta\mathcal{H} = E_{ani}S_y^2 - 2\mu_B \boldsymbol{S_y} \cdot B_y)$ (negatively) contributing to the total work, together with the major contributor $(\mathcal{H}_0 = -D_{ani}S_z^2 - 2\mu_B S_z \cdot B_z)$, although $E_{ani}/k_B = 0.046 \, K \ll |D_{ani}/k_B = -0.284 \, K]$ [25].

It is hard to consider all possible values of $(-10 \leq S_y \leq 10)$ and $(-10 \leq S_z \leq 10)$ but we may still use two full values $(S_y = \pm 10)$ and $(S_z = \pm 10)$ to estimate the range of $\frac{\delta E}{E_0}$. It is not allowed to have $(S_y = \pm 10)$ and $(S_z = \pm 10)$ simultaneously since $S_y^2 + S_z^2 = 10^2$ at any time point. As exemplified in Equation (2), the energy calculation is an integration process with a relatively long time period, rather than an instantaneous moment. Hence, we have:

$$\frac{\delta E}{E_0} = \frac{|E_{ani}S_y^2 - 2\mu_B \boldsymbol{S_y} \cdot B_y|}{|-D_{ani}S_z^2 - 2\mu_B S_z \cdot B_z|}$$

$$= \frac{|0.046K \cdot 1.38 \times 10^{-23}J \cdot K^{-1}(\pm10)^2 - 2 \times 9.27 \times 10^{-24}J \cdot T^{-1}(\pm10)(\pm1)T|}{|0.284K \cdot 1.38 \times 10^{-23}J \cdot K^{-1}(\pm10)^2 - 2 \times 9.27 \times 10^{-24}J \cdot T^{-1} \times (\pm10)(\pm0.1)T|}$$

$$= \frac{|6.35 \times 10^{-23}J \pm 1.85 \times 10^{-22}J|}{|3.92 \times 10^{-22}J \pm 1.8510 \times 10^{-23}J|} \approx (30\sim66)\%, \qquad (14)$$

in which the lower bound is close to the above measured $\frac{W_{1,3}}{W_{2,4}} = 35\%$ that will be used to adjust the calculated work.

According to our calculations, the work done by $B_z$ is $1.9 \times 10^{-23} J$, which is close to the measured work (Figure 11): $W_{2,4} = W_4 - W_2 = (1.7 \pm 0.3) \times 10^{-23} J$; the work done by $B_y$ is $35\% \times 1.9 \times 10^{-23} J = 6.7 \times 10^{-24} J$, which is close to the measured work [Figure 11(b)]: $W_{1,3} = W_3 - W_1 = (-6 \pm 2) \times 10^{-24} J$. Using the same subtraction used in the giant spin measurements [9], the total work is $1.9 \times 10^{-23} J - 6.7 \times 10^{-24} J = 1.2 \times 10^{-23} J$, which is close to the measured work: $W = |W_{2,4}| - |W_{1,3}| = (1.1 \pm 0.3) \times 10^{-23} J$ [9].

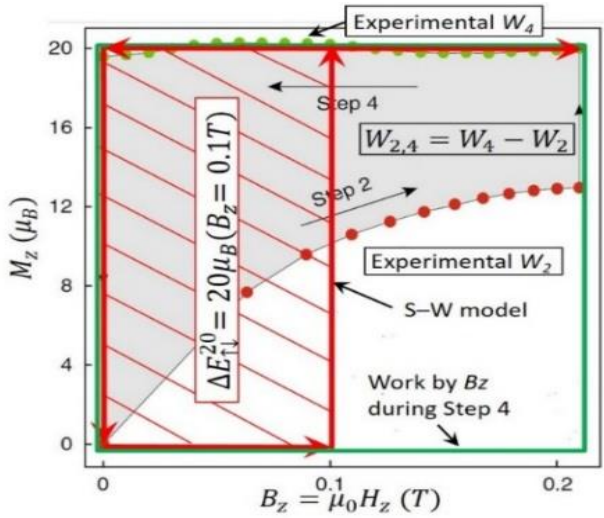

(a)  Longitudinal magnetization;

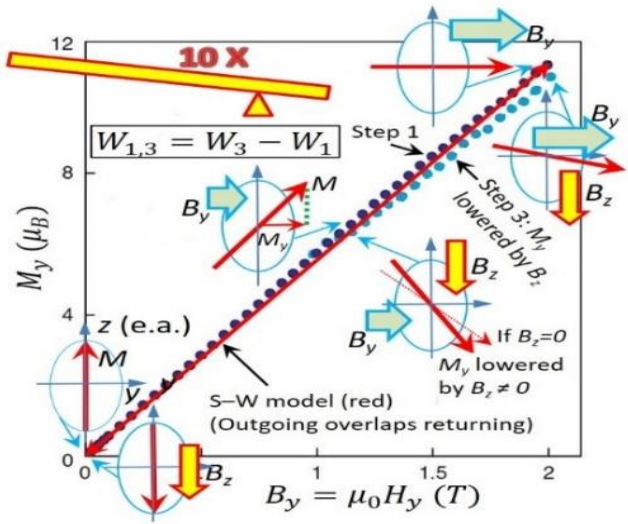

(b) transverse magnetization.

**FIGURE 11. Experimentally measured work vs. theoretically predicted work for the giant spin qubit [9]. (a) According to the Stoner–Wohlfarth (S-W) model [24],** $M_z$ **jumps to its maxima of** $20\mu_B$ **at the end of Step 2, remains constant maxima during Steps 3 & 4, and returns to zero at the end of Step 4 [the real-world hysteresis loop of a giant spin is shown in Figure 7 (middle right)]. The shaded rectangle in red** $[\Delta E_{1\uparrow}^{20} = 20\mu_B(B_z = 0.1T)$, **where** $B_z = 0.1~T$ **takes the average value of the full range** $(0\sim0.2~T)$**] corresponds to the net work done onto the switched giant spin by** $B_z$**. The energy gain (represented by the rectangle in green) in Step 4 compensates all energy cost (represented by the white area) in Step 2. The net work was measured as** $W_{2,4} = W_4 - W_2 = (1.7 \pm 0.3) \times 10^{-23}~J$**. Most importantly,** $\Delta E_{1\uparrow}^{20} = 20\mu_B B_z$ **(in red) approximates** $W_{2,4} = W_4 - W_2$ **(in grey), which successfully identifies the dominant contribution to the total work of erasure. (b)** $M_y$ **increases with increasing** $B_y$ **during Step 1 and reach its maxima of** $12\mu_B$ **at the end of Step 3, stays there during Step 2 and returns to the origin at the end of Step 3. If** $B_z = 0$**, the area enclosed by the outgoing and returning paths should be zero since the outgoing path overlaps the returning one according the S-W model. If** $B_z \neq 0$**, the returning path (Step 3) is lowered by** $B_z$ **[9]. This non-zero area was measured as** $W_{1,3} = W_3 - W_1 = (-6 \pm 2) \times 10^{-24}~J$**. The lever rule is illustrated, in which any hysteresis of** $M_y$ **measured in (b) will be 'amplified' by a factor of 10 in terms of subtracting** $M_y B_y$ **from** $M_z B_z$ **[measured in (a)] to obtain the net work.**

In the above calculations, the subtraction was fairly performed since $|W_{1,3}|$ actually came from $|W_{2,4}|$ due to the bias effect of small $B_z$ in Step 3 [Figure 11(b)], i.e., the nonzero area $|W_{1,3}|$ was originated by $B_z$ (otherwise it is zero in theory). This bias effect is similar to what was demonstrated in Figure 7 (bottom left), in which it is $B_y$ that assists the reversal by $B_z$ in terms of lowering the energy barrier.

As summarised in Table 3, Equation (12) yields $1.2 \times 10^{-26}~J$ for the single spin [7] flipped by $B_z = 1.3 \times 10^{-3}~T$, which is close to the theoretical Landauer bound ($9.6 \times 10^{-27}~J$) at the corresponding experimental temperature of 1 $mK$, $1.2 \times 10^{-23}~J$ for the giant spin [9] flipped by $B_z = 10^{-1}~T$ with the assistance of $B_y = 1~T$, which is close to the theoretical Landauer bound $(9.6 \times 10^{-24}~J)$ at the corresponding experimental temperature of 1 $K$, and $4 \times 10^{-21}~J$ for the nanomagnet [8] flipped by $B_y = 20 \times 10^{-3}~T$ [in the presence of $B_x = 20 \times 10^{-3}~T$ used to remove the uniaxial anisotropy barrier due to no tunnel through it in this classical binary system (Figure 6, bottom)], which is close to the theoretical Landauer bound $(2.9 \times 10^{-21}~J)$ at the corresponding experimental temperature of 300 $K$, respectively.

As summarised in Table 3, Equation (12) was experimentally verified by all three experiments (the single spin experiment [7], the giant spin experiment [9] and the nanomagnet experiment [8]) to universally quantify the energy required to erase an orientation-encoded bit, regardless of whether the information carrier is a single spin, a $20\mu_B$ giant spin, or a nanomagnet (containing $10^4$ spins).

TABLE 3. Comparison of three orientation-encoded binary systems.

| Info carrier | Single spin [7] | Giant spin [9] | Nanomagnet [8] |
|---|---|---|---|
| The moment size | $1\mu_B$ | $20\mu_B$ | $10^4\,\mu_B$ |
| Magnetic field $B_z$ | $B_z = 1.3 \times 10^{-3}~T$ | $B_z = 0.1~T$ | $B_x = 20$ mT $B_y = 20$ mT |
| $\Delta E_{1\uparrow}^N$ $= N\mu_B B_z$ (Eq.12) | $\Delta E_{1\uparrow} = \mu_B B_z$ $= 1.2 \times 10^{-26} J$ | $\Delta E_{1\uparrow} =$ $20\mu_B B_z$ $\times (1 - 35\%)$ $= 1.2 \times 10^{-23} J^*$ | $\Delta E_{1\uparrow} =$ $10^4\mu_B B_x$ $+10^4\mu_B B_y$ $\approx 4 \times 10^{-21} J$ |
| Landauer's Bound at T | $9.6 \times 10^{-27} J$ at T=1 mK | $9.6 \times 10^{-24} J$ at T=1 K | $2.9 \times 10^{-21} J$ at T=300 K |
| Measured work | $1.2 \times 10^{-26} J$ | $(1.1 \pm 0.3) \times 10^{-23} J$ | $\sim 4 \times 10^{-21} J$ |
| Spin relaxation time (Eq.21) | 134 s | > 100 $s$ if $B_y = 0$; $\sim 10^{-6}~s$ if $B_y = 2T$. | Seconds |
| Quantum tunnelling? | Yes | Yes | No |

*The calculated work done by the longitudinal magnetic field $B_z$ was adjusted to include the perturbation brought by the transverse magnetic field $B_y$ based on Equation (14).

Simply speaking, our method in this paper is to use the deduced formula [Equation (12)], verified by the giant spin experiment [9] and the nanomagnet experiment [8], to calculate the energy required to erase a single spin qubit extracted from the spin-spin experiment [7]. Our result is $1.2 \times 10^{-26} J$, which is slightly larger than the corresponding theoretical Landauer bound ($9.6 \times 10^{-27} J$). We view such a non-violation of Landauer's bound as an experimental verification in terms of the calculated energy being slightly larger than the bound, rather than being smaller than it or dramatically different from it. This is not a coincidence.

Energy near $k_B T \ln 2$ to erase a spin qubit at the expense of a long spin relaxation time is theoretically sensible and experimentally verified [7]. A single spin extracted from the aforementioned spin-spin magnetic interaction experiment [7] demonstrates that Landauer's bound can be quantitatively approached at an approaching rate of $\frac{1 \times 10^{-26} J}{1.2 \times 10^{-26} J} = 83\%$ via quantum spin tunnelling (Figure 12). The approaching rate is defined as the rate between Landauer's bound (at the corresponding temperature, at which an information system is operating) and the consumed energy to erase a bit of information in this information system, which describes how much the energy consumption can approach Landauer's bound. If it is 100%, Landauer's bound is reached.

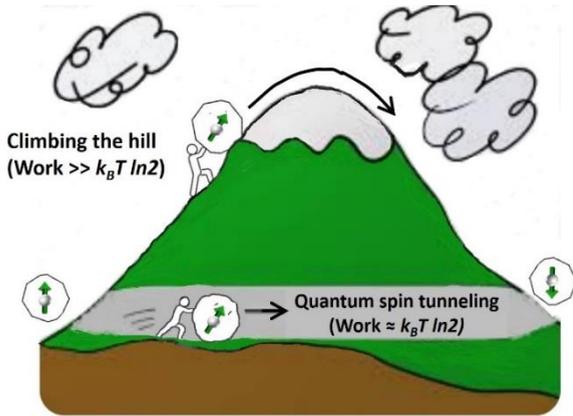

FIGURE 12. Quantum spin tunnelling penetrates the thermal energy barrier (Landauer's bound [11]) and provides an energy-efficient 'shortcut' for spin reversal, which is dramatically different from classical information manipulation (Figure 10). The cost of erasing a spin qubit does not come from "climbing the hill", but from 'tunnelling through the hill'. Quantum mechanics provides a simple answer to the question of whether it is possible to approach Landauer's bound in the envisaged extreme regimes.

Table 3 indicates that the single spin system, in which a spin is weakly bound in an ion, is superior by nature to the giant spin (molecular nanomagnet) system, in which a strong anisotropy is present related to the local structure (this strong anisotropy is caused by crystal electric field from the surrounding ligands, which is related to the local structure instead of the lattice). Although both the giant spin [9] and single spin [7] use quantum spin tunnelling, the latter is much more 'agile' than former in terms of the energy required to flip it. Note that the 'clumsiness' of the giant spin is not only due to its 'giant' size ($20\mu_B$) but also due to its strong interaction (100 times the external magnetic field in the spin-spin magnetic interaction experiment [7]) with the surroundings.

This quantum spin tunnelling effect to facilitate reversal of a qubit in quantum computing can be further analysed by rewriting the giant spin Hamiltonian in Equation (3) as:

$$\mathcal{H} = -D_{ani}S_z^2 + E_{ani}(S_x^2 - S_y^2) - 2\mu_B S_x \cdot B_x - 2\mu_B S_y \cdot B_y - 2\mu_B S_z \cdot B_z. \quad (15)$$

If $B_x = 0$ and $B_y = 0$, then the eigenvectors of this Hamiltonian are the eigenvectors $|m\rangle$ of $S_z$ and the eigenvalues of this Hamiltonian are therefore quantized and equal to

$$E_m = -D_{ani}m^2 - 2\mu_B m B_z, \quad (16)$$

where $s$ designates the eigenvalues of the spin operator $S_z$ and $m = -S_z, -(S_z - 1), -(S_z - 2), \ldots, (S_z - 1), S_z$. Thus, there are $(2S_z+1)$ eigenvectors and $(2S_z+1)$ eigenvalues.

If $B_z \neq 0$, then we obtain $[\Delta E_{1\downarrow}^N = (N\mu_B)B_z]$ [Equation (12)] that can be used to estimate the work required for erasure of each ($20\mu_B$) giant spin qubit as we did.

If $B_z = 0$, then the eigenvectors $|m\rangle$ and $|-m\rangle$ are degenerate as schematized in Figure 13(a), in which splitting of the $(2S+1)$ energy levels can be seen even in the absence of an applied magnetic field (zero-field splitting, ZFS) [25]. If $B_z \neq 0$, then it is also possible to have degeneracy ($E_m = E_{m'}$) of two eigenvectors $|m\rangle$ and $|m'\rangle$ for certain values of $H_z$ [25]. As an example, Figure 13(b) schematizes the energy levels in the double well when $E_{m=-S+1} = E_{m'=S}$ is satisfied. That is, tunnelling is possibly triggered (with the assistance of an appropriate transverse field $B_x$ or $B_y$) between the lowest (ground) state of the right-hand well and the first excited state of the left-hand well.

The value of $B_z$ can be calculated from $E_{m=-S+1} = E_{m'=S}$ with $D_{ani}/k_B = -0.294 K$ [9] as below:

$$E_{m=-S+1} = -D_{ani}S^2 + 2SD_{ani} - D_{ani} + 2\mu_B SB_z - 2\mu_B B_z, \quad (17)$$

$$E_{m'=S} = -D_{ani}S^2 - 2\mu_B SB_z, \quad (18)$$

$$B_z = -\frac{D_{ani}}{2\mu_B} = -\frac{-0.294 \, K \times 1.380 \times 10^{-23} \, J \cdot K^{-1}}{2 \times 9.274 \times 10^{-24} J \cdot T^{-1}} \approx 0.22 \, T, \quad (19)$$

which agrees well with the actual strength [$(0\sim0.2) \, T$] of $\mu_0 H_z$ used in the giant spin reversal experiment [9].

As mentioned above, if an energy gap $\Delta$ equals the tunnelling bias $\varepsilon$ (defined in Figure 8):

$$\varepsilon = [-D_{ani}(-9)^2 - 2\mu_B(-9)B_z] - [-D_{ani}(-10)^2 - 2\mu_B(-10)B_z]$$
$$= 18|D_{ani}| = 7.303 \times 10^{-23} \, J, \quad (20)$$

then tunnelling is possibly triggered for $E_{m=-9} = E_{m'=10}$.

Figure 13(b) vividly illustrates spin relaxation combining tunnelling with spin-phonon coupling (a phonon may be caused by lattice vibration [25]). Spin reversal occurs by the system absorbing phonon energy in the excitation and reaching an excited state (represented by a upward arrow in red/yellow), the spin tunnelling to the opposite side of the potential barrier, the system emitting phonon(s) in the de-excitation (represented by a downward arrow in red/yellow) and eventually reaching the ground state.

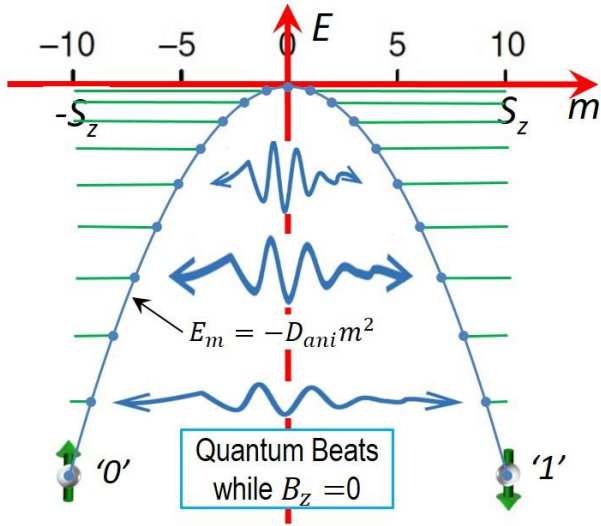

(a) Zero field splitting (ZFS);

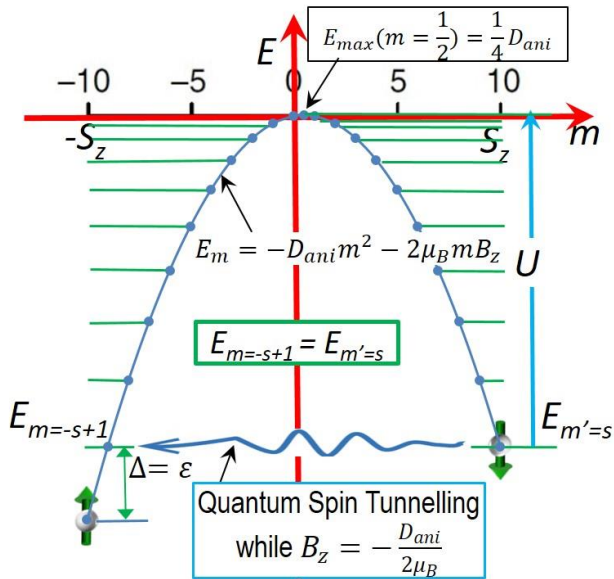

(b) subject to a longitudinal magnetic field $B_z$.

**FIGURE 13. Equation (16) as a function of $m$. (a) Zero field splitting (ZFS) with the aforementioned quantum beats (represented by three blue double arrows); (b) subject to a longitudinal magnetic field ($B_z = -\frac{D_{ani}}{2\mu_B}$) that satisfies $E_{m=-s+1} = E_{m'=s}$. With $\Delta_{m=-10} = \varepsilon$ between $m = -9$ and $m = -10$, tunnelling (represented by two blue single arrows) is possibly triggered from the lowest (ground) state of the right-hand well to the first excited state of the left-hand well. Note that (reversible) quantum beats cannot be used for qubit erasure that should be irreversible in terms of the erased information being lost permanently, whereas (irreversible) quantum spin tunnelling can. Also note that the energy landscape is a parabola ($\propto m^2$) whose maximum appears at $m = \frac{1}{2}$ in (b) since $\frac{dE_m}{dm} = -2D_{ani}m - 2\mu_B\left(B_z = -\frac{D_{ani}}{2\mu_B}\right) = (-2m+1)D_{ani} = 0$. This maximum energy is small: $E_{max}(m = \frac{1}{2}) = -D_{ani}\left(\frac{1}{2}\right)^2 - 2\mu_B\frac{1}{2}B_z = \frac{1}{4}D_{ani}$. The activation energy is $U = |E_{max}| + |E_{m'=s}| \approx |E_{m'=s}|$.**

As summarised in Table 2, this work is the first experimental verification of Landauer's bound on a single spin that is the smallest in size among various information carriers. Nevertheless, Landauer's bound has already been significantly challenged over the past two decades [26]. In 2000, Shenker argued that it is plain wrong since logical irreversibility is unrelated to heat dissipation [27]. In 2003, Bennett suggested that it is a restatement of the second law of thermodynamics [28]. In 2011 and 2019, Norton showed that the thermal fluctuation selectively neglected in the previous proofs fatally disrupted the intended operation [29][30]. Further research is needed to determine whether Landauer's bound is merely the consequence of the second law of thermodynamics [31] and whether we have to presume its demise despite the many mysteries uncovered with it over the past 60 years [17][32].

A large cooling facility that is peripheral to the central quantum processor is needed for today's few-qubit quantum computers, in which the fundamental energy given by $\Delta E = \mu_B B$ merely represents a small part of the overall energy bill. However, the cooling energy is unlikely to scale linearly with the increased number of qubits [22]; hence, its relative proportion will become less dominant in the future.

## V. NEAR-LANDAUER-BOUND ADIABATIC QUANTUM COMPUTER

As listed in Table 3, the above spin-encoded quantum computer may be slow: for the single spin [7], the spin relaxation time is 134 $s$, as shown in Figure 5; for the giant spin [9], the spin relaxation time $\tau_{rel}$ follows Arrhenius's law in the form of Landauer's bound ($k_B T$):

$$\tau_{rel} = \tau_0 exp\left(\frac{U}{k_B T}\right) = 10^{-8}\exp\left(\frac{25k_B}{k_B}\right) > 100\ s, \quad (21)$$

where $\tau_0 \approx 10^{-8}\ s$ is the attempt time, and $T = 1\ K$ for the giant spin [9]. The activation energy $U$ in the above equation was estimated [25] by Equation (18) as: $U/k_B = |E_{max}| + |E_{m'=s}| \approx |E_{m'=s}| = |-D_{ani}S^2 - 2\mu_B SB_z| \approx 25\ K$ while $|B_z| \geq \frac{D_{ani}}{2\mu_B}$, as illustrated in Figure 13(b). By applying a transverse magnetic field $B_y = 2\ T$, an effectively lowered activation energy $U/k_B \approx 5\ K$, which dramatically shortens the relaxation time to $\tau_{rel} = 10^{-8}\exp\left(\frac{5k_B}{k_B}\right) \approx 10^{-6}\ s$ [9]. The above two estimated time scales ($> 100\ s$ and $\sim 10^{-6}\ s$) agreed with the experimental measurements (71.2 $s$ and $\leq 10^{-6}\ s$) [9].

Amazingly, a slow evolution is good for adiabatic quantum computing [33][34][35]. If the Hamiltonian of a quantum computer depends on time, then its quantum state vector should satisfy the equation

$$i\hbar\frac{d|\Psi(t)\rangle}{dt} = \hat{H}(t)|\Psi(t)\rangle. \quad (22)$$

At a given time, the Hamiltonian will have a set instantaneous eigenvectors, such that $\hat{H}(t)|\Psi_n(t)\rangle =$

$E_n(t)|\Psi_n(t)\rangle$. For example, its ground state will also be time dependent.

In adiabatic quantum computing, we need to find a (potentially complicated) Hamiltonian whose ground state describes the solution to our problem of interest. First, we initialise our system with a simple Hamiltonian to the ground state (i.e., $|\Psi(0)\rangle = |\Psi_g(0)\rangle$). Next, we let the simple Hamiltonian evolve to the desired complicated Hamiltonian in an adiabatic way [the ground state of our system is always separated from the excited states by a finite energy gap (i.e., $E_1(t) - E_0(t) > 0$)]. By the adiabatic theorem [1], if the Hamiltonian evolves slowly enough to avoid the metastable states (none of which is close to our solution), then our system remains in the ground state (i.e., $|\Psi(t)\rangle = |\Psi_0(t)\rangle$), and the final state of our system encodes the solution to our problem in one go (evolution).

The adiabatic evolution in the case of a single spin qubit can be explained as follows. The time-dependent Hamiltonian

$$\hat{H}(t) = -\frac{1}{2}\begin{pmatrix} \varepsilon(t) & \Delta \\ \Delta & -\varepsilon(t) \end{pmatrix} \tag{23}$$

depends on time through the bias $\varepsilon$. Its ground state is $|\Psi_g(-\infty)\rangle = \begin{pmatrix} 0 \\ 1 \end{pmatrix}$ (i.e., in the $|\downarrow\rangle$ potential well) when $\varepsilon = -\infty$ and $|\Psi_g(\infty)\rangle = \begin{pmatrix} 1 \\ 0 \end{pmatrix}$ ($|\uparrow\rangle$ potential well) when $\varepsilon = \infty$.

As shown in Figure 14, if the qubit is initially in the ground state and the bias then slowly changes from minus to plus infinity, then the qubit will have time to tunnel from the $|\downarrow\rangle$ to $|\uparrow\rangle$ potential well, and will end up in the ground state. If the evolution is too fast by rapidly changing the bias, then the qubit will remain trapped in the $|\downarrow\rangle$ potential well, that is, in the excited state. This 'trapping' is a major source of error in adiabatic quantum computing [1][33][34][35].

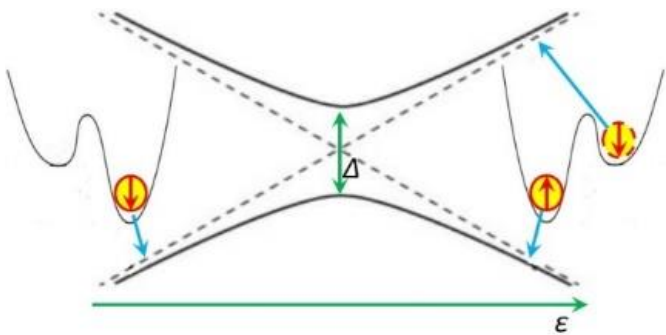

**FIGURE 14. Near-Landauer-bound adiabatic quantum computing with slow evolution of a spin qubit. If the evolution is too fast, then the qubit will jump from the ground state to the excited state by mistake [1][33][34][35].**

The near-Landauer-bound adiabatic quantum computer has some obvious advantages: any quantum program can be run on it; the quantum gates can be replaced by qubit-qubit interactions; there is no need to precisely manipulate the states of qubits; and the ground state is the lowest energy state [1]. Therefore, as long as we run the system slowly and keep the ambient temperature lower than the energy gap between the excited state and the ground state, we can suppress the noise coming from the outside and preserve the required quantum coherence for a long time.

## VI. CONCLUSIONS AND DISCUSSIONS

The question of whether it is possible to approach the classical Landauer bound in the era of quantum computing is an important one. Landauer's principle states that there is a minimum amount of energy required to perform a computation, and this limit has been a fundamental constraint on the efficiency of classical computing.

In the context of quantum computing, the situation is more complicated because quantum systems behave differently from classical systems. Quantum computers use quantum bits, or qubits, which can exist in a superposition of states and can be entangled with other qubits. This allows quantum computers to perform certain computations much more efficiently than classical computers.

However, the question of whether quantum computing can approach the Landauer bound is still an open one. Some researchers have suggested that the principles of quantum mechanics may allow for more efficient computation than classical computing, and that the Landauer bound may not apply in the same way. Others have suggested that there may be fundamental limits to the efficiency of quantum computing that are analogous to the Landauer bound in classical computing.

Overall, it is an open and active area of research to explore the fundamental limits of quantum computing and to determine whether it is possible to approach the classical Landauer bound in this fundamentally different computing paradigm.

In this study, we presented that quantum spintronics using single spin qubits may represent a novel energy-efficient near-Landauer bound computing technology. We found that although both a single spin and a giant spin use quantum spin tunnelling, the single spin is much more 'agile' than the giant spin (whose 'clumsiness' is not only due to its 'giant' size but also due to its strong coupling with the surroundings) in terms of the former consuming only 1/1,000 of the energy of erasing the latter. Overall, our study contributes to the understanding of the fundamental limits of quantum computing and the potential of quantum spintronics to provide a more energy-efficient computing paradigm.

In the (20 $\mu_B$) giant spin reversal experiment [9], the horizontal magnetic field required to flip the giant spin is large to overcome the strong magnetic anisotropy in the crystal lattice structure. The equivalent anisotropy field $H_k$ can be estimated by:

$$\mu_0 H_k = \frac{|D_{ani}|s^2}{M_s(\mu_B)} = \frac{0.294\ K \times 10^2 \times 1.380 \times 10^{-23}\ J \cdot K^{-1}}{20 \times 9.274 \times 10^{-24} J \cdot T^{-1}} \approx 2.2\ T, \tag{24}$$

where the numerator is based on Equation (16) with $H_z = 0$, and the denominator accounts for the magnetic moment ($J \cdot T^{-1}$) [9][25][36][37]. This strong anisotropy field of $2.2\ T$ establishes an upper limit to the coercivity $H_c$ as shown in Figure 7 (middle right). In practice, we have $\mu_0 H_c \approx 0.1\ T \ll \mu_0 H_k \approx 2.2\ T$ (Brown's paradox [38]) or $H_c = \alpha H_k$, in which $\alpha \ll 1$ is called the Kronmüller factor [39][40].

In fact, there are many magnetic materials with isotropic magnetic interactions (Heisenberg exchange interactions) [41]. Hopefully, there will be more and more research groups who will get closer and closer to Landauer's bound without using a strong magnetic field to erase a spin qubit in their experiments in the future.

According to Koomey's law [42][43], Landauer's bound limits irreversible operations such that the increase in the computing power efficiency (the number of computations per joule of energy dissipated) will come to a halt around 2050.

As summarized in Table 3, our current calculations are supported by at least three independent experiments within a reasonable tolerance. Furthermore, we even found that even the famous Stern-Gerlach experiment [44] may be the first demonstration of Landauer's principle, as illustrated in Figure 15.

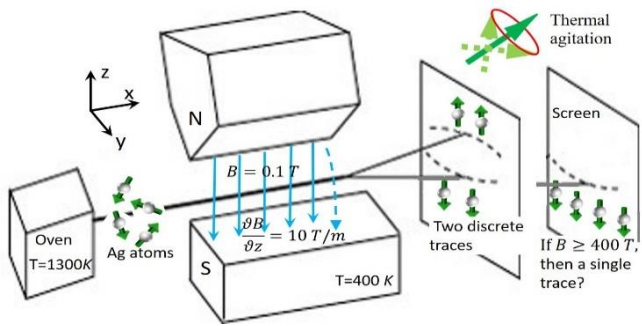

**FIGURE 15. Is the Stern-Gerlach experiment the first demonstration of Landauer's principle?** Silver atoms travel through an inhomogeneous magnetic field, being deflected up or down depending on their spin orientation. The screen reveals two discrete traces. The magnetic field has treble roles to play [45]: *B* sets the spin quantization axis and provides a torque on the spins for a potential reversal if the thermal agitation (in the form of Landauer's bound) mentioned in Figure 10 is overcome, whereas δ*B* provides a net force on the spins to deflect the atoms' trajectory. Historically, this experiment convinced physicists of spin quantization in all atomic-scale systems.

We can use Equation (12) to calculate the erasure energy:

$$\Delta E^{N=1}_{\uparrow\downarrow} = (1\mu_B)(B_z = 0.1\ T) \approx 10^{-23}\ J \cdot T^{-1} \times 0.1\ T = 10^{-24}\ J \ll k_B(T = 400\ K)\ln 2 = 1.38 \times 10^{-23} J \cdot T^{-1} \times 400\ K \times 0.69 \approx 4 \times 10^{-21}\ J. \quad (25)$$

This is why Stern and Gerlach still observed two discrete traces of the quantized spins on screen [44]. If they had increased the strength of the magnetic field $B_z$ from $0.1T$ to $\frac{4 \times 10^{-21} J}{10^{-23} J \cdot T^{-1}} = \mathbf{400\ T}$, they would have seen the collapse of the two discrete traces to one only owing to the flip of the quantized spins. As illustrated in Figure 10, for a bit of orientation-encoded information, we need to input a certain amount of energy above Landauer's bound to reach the designated erasure state from the random data state since the orientation may fluctuate to take an arbitrary direction due to thermal agitation.

## ACKNOWLEDGMENT

We thank Dr. Shlomi Kotler (the Hebrew University of Jerusalem) for discussions on his magnetic spin-spin interaction experiment and his kind permission for us to redraw the experimental setup. We thank Dr. Rocco Gaudenzi (Max-Planck Institute & Università di Verona) and Professor Fernando Luis (Universidad de Zaragoza) for providing the technical details of their giant spin reversal experiments. We also thank Prof. Francisco Javier Cao Garcia (Universidad Complutense de Madrid) and Dr. Sai Vinjanampathy (Indian Institute of Technology Bombay) for commenting our work.

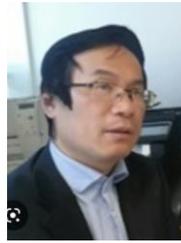

**Frank Z. Wang** (M'96–SM'00) was a Professor in Future Computing and the Head of the School of Computing (2010-2016), University of Kent, UK. Professor Wang's research interests include quantum computing, artificial intelligence, neuromorphic computing, non-Turing architecture, memristors as a new computing paradigm, unconventional computing, brain-like computer, deep learning, green computing, grid/cloud computing, and data storage & data communication, etc. He has published intensively on top journals, including IEEE Transactions on Computers, ACM Operating System Review, IEEE Transactions on Quantum Engineering, Quantum Information Processing (Springer Nature), Neural Networks, Information Sciences, IEEE Transactions on Circuits and Systems, Applied Physics Letters, IEEE Electron Device Letters, Journal of Applied Physics, and top conferences, including ACM/IEEE SuperComputing, IEEE Big Data, IEEE Mobile Services, EuroPar, IC, and IEEE Non-Volatile Memory. So far, he has attracted research fundings over £5m from the UK government and the European Commission. Professor Wang has chaired over 10 international conferences/symposiums. He is a regular keynote/invited speaker worldwide, for example at Princeton University, Carnegie Mellon University, CERN, Hong Kong University of Sci. & Tech., Tsinghua University (Taiwan), Jawaharlal Nehru University, Aristotle University, and the University of Johannesburg. In 1996, Frank designed and developed a new type of random access memory using the spin-tunnelling effect at Tohoku University, Japan. This device was the first of its kind worldwide. Frank obtained his PhD degree in the UK in 1999. Only 5 years after his PhD degree, he was appointed Chair Professor and Director of the Centre for Grid Computing at CCHPCF (Cambridge-Cranfield High Performance Computing Facility). CCHPCF is a collaborative research facility at the Universities of Cambridge and Cranfield (with an investment size of £40 million). Prof Wang and his team won an ACM/IEEE Super Computing finalist award in USA in 2007. Between 2017 and 2018, he spent his sabbatical at Tsinghua University. Prof. Wang is a Fellow of the British Computer Society. He is the Chairman (UK & Republic of Ireland Chapter) of the IEEE Computer Society. Professor Wang is also the IEEE Computer Society Region 8 Area 2 Coordinator in charge of over 10 western European countries including the UK, France, Germany, Ireland, Netherlands, Luxemburg, Belgium, Denmark, Switzerland, Spain, Portugal, Austria and Iceland. He sat on the UK Government EPSRC e-Science, Hardware for Efficient Computing and ICT Panels and the Irish Government High End Computing Panel for Science Foundation Ireland (SFI).